\documentclass[12pt,letterpaper]{article}

\usepackage[includeheadfoot,
            marginratio={1:1,2:3},
            width=412pt,
            height=688pt,]{geometry}

\usepackage{amsmath}
\usepackage{amsfonts}
\usepackage{amssymb}
\usepackage{stmaryrd}
\usepackage{graphicx}
\usepackage{empheq}
\usepackage{paralist}

\usepackage{graphicx}
\usepackage{multirow}
\usepackage{color}

\newcommand{\nc}{\newcommand}
\nc{\lb}{\llbracket}
\nc{\rb}{\rrbracket}
\nc{\gl}{\llbracket}
\nc{\gr}{\rrbracket}

\newcommand{\eq}[1]{\begin{equation}
                     \begin{split} #1 \end{split}
                     \end{equation}}

\newcommand{\ov}{\overline}
\newcommand{\defeq}{\mathrel{\mathop:}=}

\newcommand{\slnabla}{\slash\hspace{-7pt}\nabla}

\allowdisplaybreaks[2]
\numberwithin{equation}{section}

\begin{document}

\vspace*{-1.5cm}
\begin{flushright}
  {\small
  MPP-2013-152\\
  }
\end{flushright}

\vspace{1.5cm}
\begin{center}
{\LARGE
Dimensional Oxidation of Non-geometric   \\[0.2cm]
Fluxes in Type II Orientifolds}
\vspace{0.4cm}

\end{center}

\vspace{0.35cm}
\begin{center}
  Ralph Blumenhagen$^1$, Xin Gao$^{1,2}$, Daniela Herschmann$^1$ and Pramod Shukla$^1$
\end{center}

\vspace{0.1cm}
\begin{center}
\emph{$^{1}$ Max-Planck-Institut f\"ur Physik (Werner-Heisenberg-Institut), \\
   F\"ohringer Ring 6,  80805 M\"unchen, Germany  \\
\vspace{0.4cm}
$^{2}$ State Key Laboratory of Theoretical Physics, \\
Institute of Theoretical
Physics, Chinese Academy of Sciences, \\
P.O. Box 2735, Beijing 100190, China  \\[0.1cm]
}

\vspace{0.2cm}

 \vspace{0.5cm}
\end{center}

\vspace{1cm}


\begin{abstract}
Some aspects of string compactifications with 
non-geometric fluxes are revisited in the light 
of  recent progress in double field theory.
After rederi\-ving the general form of these fluxes, 
we consider the proposed flux induced  four-dimensional
effective superpotential and oxidize its induced scalar potential
to terms in a ten-dimensional action.
This analysis is performed independently  for an explicit  toroidal type IIA
and its  T-dual  type IIB orientifold.
We show in detail that the result of this  bottom-up approach is  compatible 
with the gauged supergravity
motivated  flux formulation  of the 
double field theory action in both the NS-NS and the R-R sector. 
\end{abstract}

\clearpage

\tableofcontents



\section{Introduction}
\label{sec:intro}

The application of T-duality to known string solutions with
non-trivial three-form flux in connection  with the possible  gaugings
in gauged supergravity has led to the proposal for the existence
of non-geometric fluxes in string theory.
During the last couple of years, both their
string model building implications as well as their mathematical
description  have been under intense study.  
This revealed  a surprisingly 
rich structure opening  new physical possibilities 
and connections to current areas of mathematics.

Concretely, successively applying the Buscher rules to a flat three-dimensional 
background with constant $H$-flux,
leads in the first step to a twisted torus with a constant 
geometric $F$-flux \cite{Kachru:2002sk} and 
in the second step
\cite{Hellerman:2002ax,Dabholkar:2002sy,Hull:2004in},  
to the non-geometric $Q$-flux. Such backgrounds   can be considered
to be locally geometric, where the transition functions
between two overlapping charts
are stringy T-duality transformations. Therefore, such 
backgrounds have been called T-folds \cite{Hull:2006va}.
A final  T-duality in 
the  non-isometric third direction was conjectured to lead to 
the $R$-flux \cite{Shelton:2005cf}, for which the background is expected to be 
not even locally geometric.

Since a Kalb-Ramond three-form
supported on a Calabi-Yau threefold was  known to generate an effective
four-dimensional  superpotential, the form of the latter for the non-geometric
$Q$- and $R$-fluxes could be conjectured \cite{Shelton:2005cf}. 
Ana\-logously, one could 
derive that a combination of Ramond-Ramond fluxes and non-geometric fluxes
contribute to the R-R tadpole cancellation conditions. In fact, it was argued
that these contributions can weaken these constraints in the sense that 
their effect is like that of anti-branes
\cite{Camara:2005dc,Aldazabal:2006up}.
Turning on those more general fluxes break the no-scale structure of the
simplest type IIB superpotential and the presence of 
new stable AdS and Minkowski vacua has been shown for toroidal
examples. Clearly, it is one of the  crucial questions of string
phenomenology,  whether the flux induced scalar potential (in connection
with  instanton and D-term contributions) can stabilize all
moduli in a stable or meta-stable de Sitter minimum.
Including the non-geometric fluxes 
this question was approached
lately in \cite{deCarlos:2009qm,Danielsson:2012by,Blaback:2013ht,Damian:2013dq,Damian:2013dwa}.

It was pointed out quite early that the stringy fluxes are expected
to be closely related
to the possible gaugings in gauged four-dimensional supergravity
\cite{Derendinger:2004jn, Derendinger:2005ph, Dall'Agata:2009gv,
  Aldazabal:2011yz, Aldazabal:2011nj,Geissbuhler:2011mx,Grana:2012rr,Dibitetto:2012rk}.
As a consequence, the induced scalar potentials in four-dimensions should
match. Evidence for this was presented in the literature
\cite{Villadoro:2005cu, Aldazabal:2008zza,Aldazabal:2010ef,Dibitetto:2011gm} for
the concrete case of a $\mathbb Z_2\times \mathbb Z_2$ orbifold.
In this paper, we will revisit this  background with all NS-NS fluxes
$H,F,Q, R$ and all R-R fluxes turned on  in a type IIA and a type IIB 
orientifold  
and analyze it in very much detail, but with  special emphasis on
their  ten-dimensional origin. 

It is quite remarkable  that this entire four dimensional  analysis
could be performed without having an honest ten-dimensional framework,
which from first principles implements the non-geometric sector of 
string theory. The various types of fluxes were treated as essentially
independent objects, only constrained by a couple of Bianchi identities.
Since the number of degrees of freedom of a string is fixed, one would 
expect that there are  additional constraints, which become visible
in a just mentioned microscopic description of all the fluxes.

During the last couple of years, the underlying mathematical principles
of non-geometry were under intense investigation.
Here the main candidates are
genera\-lized geometry
\cite{Hitchin:2004ut,Gualtieri:2003dx,Grana:2008yw,Coimbra:2011nw} 
and double field theory (DFT) \cite{Hull:2009mi,Hull:2009zb,Hohm:2010jy,Hohm:2010pp,Aldazabal:2011nj}. In the former, the 
$B$-field gauge transformations and the diffeomorphisms are unified 
in $O(D,D)$ transformations acting on the generalized bundle $TM\oplus T^*M$
(see \cite{Hull:2007zu,Pacheco:2008ps,Berman:2010is} for a generalization to M-theory).
In doing so, the theory automatically  contains  $\beta$-transforms,
which, loosely speaking, open  the door to (part of) the non-geometric regime.
In fact, by performing $O(D,D)$ field redefinitions, ten-dimensional
actions involving the non-geometric $Q$ and $R$-fluxes could be written down
\cite{Andriot:2011uh, Andriot:2012wx, Andriot:2012an, Blumenhagen:2012nk,Blumenhagen:2012nt}.
Here, the differential geometry of Lie-algebroids turned out to be the relevant 
mathematical structure.

However, the detailed analysis of \cite{Blumenhagen:2013aia} led the authors
to the conclusion 
that these redefined actions do not inherently allow to
describe non-geometric $Q$- and $R$-flux string backgrounds.
The reason behind that is that the new actions only feature the
redefined symmetries of the original supergravity action, i.e.
diffeomorphisms and $B$-field gauge transformations, but for instance
no local $\beta$-transforms. 
The latter are incorporated  into DFT so that  this seems to be the 
more appropriate  framework for a global description of non-geometric backgrounds
(for  pedagogical reviews of DFT please  see \cite{Aldazabal:2013sca, Berman:2013eva}). 

In DFT not only the dimension of the bundle
is doubled but even the dimension of the underlying manifold.
Motivated by the decoupling of left- and right-movers on the string
world-sheet, besides the ordinary coordinates, one introduces 
winding coordinates. The latter are the conjugate variables to the
winding of the string. In DFT, T-duality acts by exchanging 
normal with winding coordinates and also allows to perform a
T-duality in non-isometric directions.
Motivated by string field theory, a $2D$-dimensional
action admitting global $O(D,D)$ invariance as well as invariance
under both ordinary diffeomorphisms and winding diffeomorphisms was
constructed. Due to the unphysical doubling of the coordinates, this action is
supplemented by the  strong constraints, which implements
the stringy level-matching condition on the level of the effective theory.
Thus, one way of thinking about the DFT action is 
that, in constructing this (leading order) effective string action, 
one imposes
the level-matching condition after computing the string scattering amplitudes.
 
In this paper, now equipped with  the DFT framework, 
we revisit some aspects of the 
early approaches to non-geometric fluxes. First of all, generalizing
the computation done in \cite{Blumenhagen:2012pc} to DFT,
we rederive the most general form of the $H,F,Q,R$-fluxes  in terms of
the generalized metric from  the closure of the algebra of certain differential
operators. The so obtained form is consistent with the earlier
results presented in \cite{Aldazabal:2010ef,Aldazabal:2011nj,Geissbuhler:2011mx}.
The Jacobi identities of this Lie-algebra imply 
the general form of the Bianchi identities, which are also consistent
with the recent results reported in \cite{Geissbuhler:2013uka}. 

Second and this should be considered  as the main objective of this paper, 
we perform
a closer investigation of the T-duality motivated  form of the flux induced
superpotential \cite{Shelton:2005cf,Aldazabal:2006up}. 
For that purpose, on a simple toroidal orientifold with all
invariant  geometric and non-geometric NS-NS and R-R fluxes turned on, 
we   present the computation of  the
induced scalar potential in very much detail.
This analysis is carried out for a pair of T-dual 
type IIA and type IIB orientifolds separately. Note that
this computation turns out  to be highly non-trivial, as the Bianchi identities 
of the fluxes have to be invoked many times. 
Consistent with general expectation,  
the induced scalar potential is that of a half-maximally
gauged supergravity theory.

Such a detailed comparison 
between the scalar potential
induced by the fluxed superpotential and the scalar potential of gauged 
supergravity was described   for a subset of fluxes, namely geometric and $H$-flux, in
type IIA in \cite{Villadoro:2005cu}.
Beyond that, in
\cite{Aldazabal:2008zza} is was very briefly stated at the very end of the paper
that the authors have explicitly verified the matching for  all
orientifold even fluxes in the type IIB case. Similarly such a result was
briefly stated for the special case of an isotropic $T^6$ 
in  \cite{Dibitetto:2011gm}. 
\footnote{Similar to \cite{Villadoro:2005cu},
  M-theory compactification on $G_2$ manifolds with the inclusion of
  non-geometric F-fluxes have been studied in  \cite{Dall'Agata:2005fm}.}
Our detailed computation  verifies these claims for the type IIB case 
and provides 
many  interesting details of the general computation in particular for the
less studied type IIA case. 

Next we  oxidize the four-dimensional scalar potential to an
underlying ten-dimensional action. 
We emphasize that in determining the ten-dimensional origin of 
the scalar potential, one has to keep in mind that in writing
down a superpotential, one is treating the effects of the background fluxes
as small perturbations around the Calabi-Yau geometry. 
As the main result of this paper, we find that, both in the NS-NS and in the R-R sector, 
the resulting oxidized ten-dimensional action
is compatible with the flux formulation  of the DFT action 
\cite{Aldazabal:2011nj,Geissbuhler:2011mx,Grana:2012rr,Berman:2012uy,Berman:2013cli}.
Therefore, our detailed and explicit computation can be considered to provide 
another  explicit verification of all the four and ten-dimensional
concepts developed during the last years to describe non-geometric fluxes.

\section{Preliminaries}

In this section, both for self-consistency and for formulating the
problem,
 we review a couple of relevant issues about non-geometric
fluxes. Here we essentially follow the historical development.

\subsection{T-duality}

One of the most distinctive features of string theory  is certainly
T-duality. Applying this transformation to configurations which are already
well understood has led to substantial new  insights about string theory.
Most recently,  applying T-duality to closed-string backgrounds
with non-vanishing three-form flux, revealed configurations
transcending the usual large volume geometric framework. Thus,
they have been called non-geometric.

As usual, one defines the  NS-NS two-form $B={1\over 2} B_{ij}\, dx^i\wedge dx^j$
and its field strength  $H=dB={1\over 3!} H_{ijk} \, dx^i\wedge dx^j\wedge
dx^k$, i.e. $H_{ijk}=3\partial_{[\underline{i}} B_{\underline{jk}]}$
\footnote{The
anti-symmetrization of $n$ indices
is defined with the inclusion of a prefactor $1/n!$.}.
Now, one considers  a flat three-dimensional background with
constant $H$-flux, $H_{123}=h$. Using the gauge symmetry of the $B$-field,
one can choose $B_{12}=h\, x_3$. The directions $x_1$ and $x_2$ are
isometries so that one can employ the Buscher rules for a T-duality
along these circle directions. As has been shown in detail in
\cite{Kachru:2002sk, Shelton:2005cf}
a T-duality along $x_1$ leads to a twisted three-torus which is
characterized by a geometric flux $F^1{}_{23}=h$.
A further T-duality along the isometry direction $x_2$ leads
to a non-standard type of string background, where the
transition function between two charts is given by a stringy
T-duality transformation. This non-geometric twist is
characterized by a  $Q$-flux, $Q_3{}^{12}=h$.
It now seems tempting to also pull the last index up by performing
a T-duality in the non-isometry direction $x_3$. Clearly, the
Buscher rules cannot be applied in this case, but it was
speculated that one gets a genuine non-geometric background
with constant $R$-flux, $R^{123}=h$. These backgrounds
are expected to be  not even locally geometric.
This chain of  transformations can be summarized by
\eq{
  H_{ijk} \;\xleftrightarrow{\;\; T_{k}\;\;}\;
   F_{ij}{}^{k} \;\xleftrightarrow{\;\; T_{j}\;\;}\;
  Q_{i}{}^{jk} \;\xleftrightarrow{\;\; T_{i}\;\;}\;
  R^{ijk} \; .
}

As we will review in section \ref{sec_dft}, a framework
to also describe this last T-duality in a non-isometry direction
is given by DFT. Here one formally  introduces
the canonical conjugate variable $\tilde x_i$ to the
winding operator. Thus, the space is doubled and parametrized
by standard and  winding coordinates $X^M=(\tilde x_i,x^i)$.
A T-duality in a certain direction also exchanges the corresponding normal and
winding coordinates.


\subsection{Flux induced superpotential}

Turning on fluxes on  a given string background, like e.g.
a torus or a Calabi-Yau manifold, induces  various effects.
First, the fluxes contribute non-trivially to the ten-dimensional
equations of motion and to the supersymmetry variations.
Therefore, generically supersymmetry is broken and the
equations of motion are not satisfied any longer.
This manifests itself in the four-dimensional effective
supergravity theory by an induced scalar potential, which at the level
of second order in derivatives arises from a flux induced
superpotential or Fayet-Iliopolous term.
Just working in the
four-dimensional effective theory, the idea is that the
potentially new minima of the scalar potential indeed
correspond to new true solutions to the ten-dimensional
equations of motion, in which the backreaction of the fluxes
is taken into account. To show this in detail is however a highly
non-trivial issue.

Second, the flux induces not only the just described
tadpoles for the dilaton and
the graviton, but also for certain  Ramond-Ramond p-form gauge fields.
Therefore, generically tadpole cancellation requires the
introduction of orientifold planes and D-branes in the background.

In this paper we are concerned with the effect of turning on all
kinds of geometric and non-geometric fluxes, which can best be
seen in type IIA orientifolds.
Recall that the string-frame ten-dimensional action for
the bosonic fields of the type IIA  supergravity is
\begin{eqnarray}
\label{actiontypea}
&&\hspace{-0.5cm}S_{\rm IIA}=\frac1{2\tilde\kappa_{10}^2}\int d^{10}x\ \sqrt{-G}\Big[e^{-2\Phi}\biggl(R+4(\nabla\Phi)^2
-{1\over 2}|H|^2\biggr)-{1\over 2}|G^{(2)}|^2-{1\over 2}|G^{(4)}|^2\Big]\nonumber\\
&&\qquad-\frac1{4\tilde\kappa_{10}^2}\int B\wedge dC^{(3)}\wedge dC^{(3)}\,
\end{eqnarray}
with gravitational coupling
$\tilde\kappa_{10}^2 = \frac{1}{4\pi}(4\pi^2\alpha')^4$, which we set to one
in the following.
Here we used the notation
\eq{
\label{formsquareabs}
|G^{(p)}|^2={1\over p!}G_{M_1\dots M_p}G^{M_1\dots M_p}\; .}
Note that the field strengths $G^{(p)}$ in the R-R sector also involve the
R-R-potentials of lower  degree and also the NS-NS two-form $B$:
\eq{\label{IIBRR}
G^{(2)} =d C^{(1)}\ , \quad G^{(4)}=d C^{(3)}-C^{(1)}\wedge H\, .
}
This leads to correction terms for the usual Bianchi
identities, as  for instance one gets  $dG^{(4)}+H\wedge G^{(2)}=0$.
Introducing  also a $G^{(0)}$ field strength leads to massive
type IIA. The other field strengths receive further corrections
\eq{
G^{(2)} &=d C^{(1)}-G^{(0)}\, B\, , \qquad G^{(4)}=d C^{(3)}-B\wedge G^{(2)} -{1\over 2} B\wedge B G^{(0)}\\
G^{(6)}&=d C^{(5)}-B\wedge G^{(4)} -{1\over 2} B\wedge B\wedge G^{(2)}-{1\over
  6} B\wedge B\wedge B\, G^{(0)}\, ,
}
and the Bianchi identities can compactly be written as
\eq{
\label{bianchimassive}
              d(e^B G)=0 \Leftrightarrow  dG^{(p)}+ H\wedge G^{(p-2)}=0\, .
}
Therefore, it is the combination  $e^B G$ which is closed in cohomology.
We call its background value $\ov{G}$.

In ${\cal N}=1$ supersymmetric theories in four dimensions
the scalar potential can be written as a sum of two terms,
an F-term and a positive semi-definite D-term. The former can
be derived from a holomorphic
superpotential $W$ and  the real K\"ahler potential $K$ via
\eq{\label{VF}
V_{W}=e^{K}\Big(G^{i\bar\jmath}D_i W\, D_{\bar\jmath} \ov W-3
|W|^2\Big)\,.}
Generalizing the celebrated  type IIB computation of 
Taylor/Vafa \cite{Taylor:1999ii},
the superpotential for turning on the R-R fluxes and the H-flux
in massive type IIA orientifolds on Calabi-Yau threefolds
was derived via dimensional reduction
of \eqref{actiontypea} in  \cite{Grimm:2004ua} (see also \cite{Benmachiche:2006df})
and reads
\eq{\label{gvw2a}
W={1\over 4}\, \left( -i \int_{X}
  \ov{H}\wedge \Omega^C +\int_{X} e^{iJ_c} \wedge \ov{G}  \right)\,,
}
where
the complexified K\"ahler modulus is defined
as $J_c=J+i B$, and the complex structure moduli are encoded in
\eq{
\Omega^c&={\rm Re} (i\, e^{-\phi}\, \Omega_3) + i C^{(3)} \, .
}
Thus, in type IIA the superpotential depends also
on the K\"ahler moduli, but
still no terms mixing the
complex structure and the K\"ahler moduli appear.

Applying successive T-duality for a toroidal background, it was argued in
\cite{Shelton:2005cf} (see also \cite{Aldazabal:2006up}) 
that  such mixing terms are generated by the
T-dual geometric and non-geometric fluxes. The proposed
superpotential reads
\eq{
\label{QRsuperpot}
W={1\over 4}\, \bigg( &-i\int_{X}
\ov{\mathfrak{H}}{}^C\wedge \Omega^C + \int_{X} e^{iJ_c} \wedge \ov{G}
 \bigg)\,,
}
where the background three-form flux
\eq{   \ov{\mathfrak{H}}{}^C= {1\over 6} \ov{\mathfrak{H}}{}^C_{ijk} \, dx^i\wedge dx^j\wedge dx^k
}
is defined as
\eq{
\ov{\mathfrak{H}}{}^C_{ijk}=\ov{H}_{ijk}+3\, \ov{F}^m{}_{[\underline{ij}}\,
       (-iJ_c)_{m\underline{k}]}
         &+3\, \ov{Q}_{[\underline{i}}{}^{mn} (-iJ_c)_{m\underline{j}}\,
           (-iJ_c)_{n\underline{k}]}\\[0.1cm]
    &+\ov{R}^{mnp}  (-iJ_c)_{m[\underline{i}}\, (-iJ_c)_{n\underline{j}}\,
           (-iJ_c)_{p\underline{k}]}\, .
}
As indicated, here the fluxes $\ov H,\ov F,\ov Q$ and $\ov R$ 
receive some background values, whereas
$J_c$ are still considered as the moduli of the Calabi-Yau threefold.
This means that turning on  geometric or non-geometric fluxes 
is treated as a perturbation around  the unfluxed Calabi-Yau geometry.

Via the supergravity relation \eqref{VF} the superpotential \eqref{QRsuperpot}
induces a scalar potential involving all closed string moduli.
Considering this scalar potential to be generated by the dimensional
reduction of a ten-dimensional effective action, it is   a natural
question to ask how such a ten-dimensional action involving also
the non-geometric fluxes must look like. Thus, we want  to oxidize
the four-dimensional effective action to a ten-dimensional one.

\subsection{Double field theory}
\label{sec_dft}

In order to inherently describe non-geometric string backgrounds,
where e.g. the transition functions are T-duality transformations,
one needs an effective string action which is manifestly invariant
under the full $O(D,D)$ group. The search for such an  action has led to
DFT \cite{Hohm:2010pp,Hull:2009mi,Hull:2009zb,Hohm:2010jy}, which
so far is only understood 
at leading order in a derivative expansion.


As already mentioned, the main new feature of DFT is that
one doubles the number of coordinates
by introducing   winding coordinates $\tilde x_i$ and arranges
them into a doubled vector $X^I=(\tilde x_i,x^i)$.
One defines an $O(D,D)$ invariant metric
\eq{
 \eta_{IJ}=
    \left(\begin{matrix}  0 &  \delta^i{}_j \\
            \delta_i{}^j & 0 \end{matrix}\right) \, .
}
Moreover, the dynamical fields $G_{ab}$ and $B_{ab}$  are combined
in the generalized metric
\eq{
\label{genmetric}
    {\cal H}_{IJ}=
      \left(\begin{matrix}  G^{ij} &  -G^{ik}B_{kj} \\
            B_{ik}G^{kj} & G_{ij} -B_{ik}G^{kl}B_{lj} \end{matrix}\right) \, .
}
Indices are pulled up and down with $\eta$, like for instance
\eq{
         {\cal H}^{IJ}=  \eta^{II'}\, {\cal H}_{I'J'}\, \eta^{J'J}\, .
}   
As in ordinary differential  geometry, one can introduce an $O(D,D)$
generalized non-holonomic frame via 
\eq{
     {\cal H}_{IJ}=E^A{}_I\, S_{AB}\, E^B{}_J
}
where the diagonal matrix $S_{AB}$ is defined as
\eq{
            S_{AB}=
 \left(\begin{matrix}  s^{ab} &  0 \\
            0 & s_{ab} \end{matrix}\right) \, 
}
with $s_{ab}$ being the flat $D$-dimensional Minkowski metric.
For the parametrization \eqref{genmetric} of the generalized metric one finds
\eq{
\label{gennonhol}
       E^A{}_I=
 \left(\begin{matrix}  e_a{}^i &  -e_a{}^k\, B_{ki} \\
            0 & e^a{}_i \end{matrix}\right) \, 
}
with $e_a{}^i s^{ab} e_b{}^j=G^{ij}$.

The full DFT action in $2D$ dimensions can then be written 
in terms of the ge\-neralized metric as
\begin{eqnarray}
 \label{dftaction}
  && S_{{\rm DFT}}={1\over 2}
\int
d^{2D} X\,  \; e^{-2d}\, \biggl( \tfrac{1}{ 8} \hspace{1pt}{\cal H}^{IJ} (\partial_I
     {\cal H}^{KL}) (\partial_J {\cal H}_{KL}) \\
   &&\hspace{22.5pt}  -\tfrac{1}{ 2} \hspace{1pt}{\cal H}^{IJ} (\partial_J
     {\cal H}^{KL})( \partial_L {\cal H}_{IK})
     -2 \hspace{1pt}(\partial_I d) (\partial_J {\cal H}^{IJ})
   + 4\hspace{1pt} {\cal H}^{IJ}  (\partial_I d)  (\partial_J d) \biggr) .\nonumber
\end{eqnarray}
Note that here $d^{2D} X=d^D x d^D\tilde x$, $\partial_I=(\tilde\partial^i,\partial_i)$, and $d$ denotes the
dilaton which is defined as
$\exp(-2d)=\sqrt{-|G|} \exp(-2\phi)$.
This action has been determined by invoking a number of symmetries:
First it was required to be invariant under local diffeomorphisms of the
coordinates $X^I$, i.e.\ $(\tilde x_i,x^i)\to (\tilde
x_i+\tilde\xi_i(X),x^i+\xi^i(X))$\, \footnote{
The $x^i$ dependence of these two diffeomorphisms includes  both
standard diffeomorphisms and $B$-field gauge transformations. Note that the
winding coordinate dependence of $\xi^i$ also gives what one might call
$\beta$-field gauge transformations.}. Second, the action is invariant
under a global or rigid $O(D,D)$ symmetry, which acts as
\eq{
\label{dft_trafo}
         {\cal H}'&=h^t\, {\cal H} h \,, \qquad\,  d'=d\;,\\
         X'&=h X\,, \qquad\quad \partial'=(h^t)^{-1} \,\partial \;
}
with
\eq{
            h=\left(\begin{matrix}  a & b\\
                              c & d \end{matrix}\right).
}
For  manifest $O(D,D)$ invariance and for closure of the algebra 
of infinitesimal diffeomorphisms, this action has to be
supplemented by the  strong constraint
\eq{
  \label{strong_c}
      \partial_i A\, \tilde\partial^i B +  \tilde\partial^i A\, \partial_i B=0\, .
}
Solving \eqref{strong_c} via $\tilde\partial^i=0$, the double field theory action
reduces to  the familiar action in the geometric frame. 

Besides the form of the DFT action \eqref{dftaction}, there exist equivalent
ones, which differ by terms that  are either total derivatives or are
vanishing due to the strong constraint. As will become
important later, there exist the so-called flux formulation of the DFT action,
which is motivated by the scalar potential in gauged supergravity
\cite{Aldazabal:2011nj,Geissbuhler:2011mx,Grana:2012rr} and which, as
shown in \cite{Hohm:2010xe}, is also related to the early work of 
W.Siegel \cite{Siegel:1993xq,Siegel:1993th}.

One can also parametrize the generalized metric  via
\eq{
\label{nongeomframe}
    {\cal H}_{IJ}=
      \left(\begin{matrix} g^{ij}-\beta^{ik} g_{kl}\beta^{lj} &  \beta^{ik}  g_{kj}\\
                  - g_{ik} \beta^{kj} & g_{ij}\end{matrix}\right)
}
where $\beta={1\over 2}\beta^{ab}\,\partial_a\wedge \partial_b$ denotes
an anti-symmetric bi-vector.
The geometric frame \eqref{genmetric} and this non-geometric one
are  related via the field redefinition
\eq{
            g&=G-BG^{-1}B\\
            \beta &=- g^{-1} B G^{-1}
}
which is reminiscent of the Buscher rules. The reduction of the DFT
action for this $(g,\beta)$ frame with $\tilde\partial^i=0$ has
been carried out in detail in \cite{Andriot:2012wx, Andriot:2012an}.

\subsection{Fluxes and Bianchi identities in DFT}

The form of the four kinds of fluxes $H,F,Q$ and $R$  in DFT
 was determined in \cite{Aldazabal:2010ef,Aldazabal:2011nj,Geissbuhler:2011mx}.
Here, we rederive them from the  generalization of 
the Roytenberg algebra from \cite{Halmagyi:2009te, Blumenhagen:2012pc}
to DFT.

The observation is that the fluxes $H_{abc},F^a{}_{bc},Q_a{}^{bc}$ and
$R^{abc}$  appear as
structure ``constants'' in the Roytenberg algebra \cite{Roytenberg:01}
on $TX\oplus T^*X$
\eq{
\label{commalgebra}
[e_a,e_b]&= F^c{}_{ab}{}\,
        e_c + H_{abc} \, e^c \, ,\\
[e_a,e^b]&= Q_{a}{}^{bc}\,
        e_c - F^b{}_{ac}{} \, e^c \, ,\\
[e^a,e^b]&= R^{abc}\,
        e_c +  Q_c{}^{ab} \, e^c \, .
}
A representation of this algebra can be given by
the Lie-bracket of certain vector fields on the tangent bundle
of the doubled geometry of DFT.
Then, the  Bianchi identities
for these fluxes arise from the Jacobi-identities of this
Lie-algebra.
Generalizing the computation of \cite{Blumenhagen:2012pc}, we consider
the following two  $B,\beta$-twisted  derivative operators
\footnote{Note that these twisted 
derivatives are related to the usual ones via a choice of 
generalized non-holonomic frame 
\eq{
\label{gennonhol2}
       E^A{}_I=
 \left(\begin{matrix}  e_a{}^i &  e_a{}^k\, B_{ki} \\
            e^a{}_k \beta^{ki}  & e^a{}_i+e^a{}_k \beta^{kl} B_{li}  \end{matrix}\right) \, ,
}
which contains both $B$ and $\beta$. There is a change of signs relative to 
\eqref{gennonhol}, but we intend here to use the same convention for the fluxes
as in \cite{Geissbuhler:2013uka}. This slight inconsistency does not affect
the computation in section \ref{sec_three}. Note that in \eqref{twistedderiv}
and therefore also in \eqref{gennonhol2}
we have  made an asymmetric choice in the definition of 
${\cal D}_a$ and $\tilde{\cal D}^a$.
}.
\eq{
\label{twistedderiv}
         {\cal D}_a=\partial_a+ B_{am} \tilde\partial^m\, ,\qquad\quad
    \tilde{\cal D}^a=\tilde\partial^a+\beta^{am}{\cal D}_m
}
where we have used a  non-holonomic basis
\eq{
            \partial_a=e_a{}^i \partial_i\,,  \qquad
           \tilde\partial^a=e_i{}^a \tilde\partial^i
}
with $e_i{}^a\, e_{ja}=g_{ij}$ and $e_a{}^i e_i{}^b=\delta_a{}^b$. Here we allow the $e_a{}^i$ to depend
on both normal and winding coordinates.
For the commutator of two partial derivatives one gets
\eq{
              [\partial_a,\partial_b]=f^c{}_{ab}\, \partial_c
}
with
\eq{
         f^c{}_{ab}\defeq e_i{}^c \Big( \partial_a e_b{}^i - \partial_b
         e_a{}^i\Big)\; .
}
Analogously, for the partial winding derivatives one finds
\eq{
[\tilde\partial^a,\tilde\partial^b]=\tilde{f}_{c}{}^{ab}\, \tilde\partial^c
}
with
\eq{
       \tilde f_a{}^{bc}\defeq e_a{}^i\,  \Big( \tilde\partial^b e_i{}^c -
               \tilde\partial^c e_i{}^b \Big)\; .
}
It is now a tedious, though straightforward computation to derive
the commutator algebra of the two twisted derivatives \eqref{twistedderiv}
\eq{
\label{commalgebra}
[{\cal D}_a,{\cal D}_b]&= F^c{}_{ab}{}\,
        {\cal D}_c + H_{abc} \, \tilde{\cal D}^c \, ,\\
[{\cal D}_a,\tilde {\cal D}^b]&= Q_{a}{}^{bc}\,
        {\cal D}_c - F^b{}_{ac}{} \, \tilde{\cal D}^c \, ,\\
[\tilde{\cal D}^a,\tilde {\cal D}^b]&= R^{abc}\,
        {\cal D}_c +  Q_c{}^{ab} \, \tilde{\cal D}^c
}
with the following definitions of the $H$-flux
\eq{
      H_{abc}\defeq 3\Big(\partial_{[\underline{a}} B_{\underline{bc}]}
                    + f^m{}_{[\underline{ab}}\,
                    B_{\underline{c}]m}\Big)+
               3\Big( B_{[\underline{a}m}\tilde{\partial}^m B_{\underline{bc}]}
             +B_{[\underline{a}m} B_{\underline{b}n} \tilde{f}_{\underline{c}]}{}^{mn}\Big)\, ,
}
the geometric flux
\eq{
\label{fluxgeom}
            F^c{}_{ab}\defeq f^c{}_{ab} + \tilde{\partial}^c B_{ab} + \tilde{f}_a{}^{cm}B_{mb}
               + \tilde{f}_b{}^{cm}B_{am} + \beta^{cm} H_{mab} \, ,
}
the non-geometric $Q$-flux
\eq{
\label{exqflux}
            Q_c{}^{ab}\defeq &\tilde f_c{}^{ab} + {\partial}_c \beta^{ab} + {f}^a{}_{cm} \beta^{mb}
               + {f}^b{}_{cm}\beta^{am}  \\
   & +B_{cm} \tilde\partial^m \beta^{ab} + 2 \beta^{m[\underline{a}} \tilde\partial^{\underline{b}]} B_{mc}
   +2B_{cm} \tilde f_n{}^{m[\underline{a}} \beta^{\underline{b}]n}
  +2 \beta^{m[\underline{a}}  \tilde f_c{}^{\underline{b}]n} B_{mn}\\
&+ \beta^{am}\beta^{bn} H_{mnc}
}
and the non-geometric $R$-flux
\eq{
\label{fluxr}
  R^{abc}\defeq\, &3\Big(\tilde\partial^{[\underline{a}} \beta^{\underline{bc}]}
                    + \tilde f_m{}^{[\underline{ab}}\,
                    \beta^{\underline{c}]m}\Big)+
               3\Big( \beta^{[\underline{a}m} {\partial}_m \beta^{\underline{bc}]}
             +\beta^{[\underline{a}m} \beta^{\underline{b}n}  {f}^{\underline{c}]}{}_{mn}\Big)\\    &  +3\Big( B_{mn} \beta^{[\underline{a}m} \tilde{\partial}^n \beta^{\underline{bc}]}
     +\beta^{[\underline{a}m} \beta^{\underline{b}n} \tilde\partial^{\underline{c}]} B_{mn}+
2\beta^{[\underline{a}m} \beta^{\underline{b}n}  \tilde{f}_{[\underline{m}}{}^{\underline{c}]k} B_{k\underline{n}]}\Big)\\[0.1cm]
&+ \beta^{am}\beta^{bn}\beta^{cp}  H_{mnp} \, .
}
Here we have used the strong constraint \eqref{strong_c}.
We observe  that, even in a holonomic frame, all four types of fluxes
receive contributions.  We have the $H$-flux
\eq{
      H_{ijk}\defeq 3 \partial_{[\underline{i}} B_{\underline{jk}]} +
               3  B_{[\underline{i}m}\tilde{\partial}^m B_{\underline{jk}]} \, ,
}
the geometric flux
\eq{
            F^k{}_{ij}\defeq \tilde{\partial}^k B_{ij} + \beta^{km} H_{mij} \, ,
}
the non-geometric $Q$-flux
\eq{
\label{exqflux2}
            Q_k{}^{ij}\defeq & {\partial}_k \beta^{ij}  +B_{km} \tilde\partial^m \beta^{ij} + 2 \beta^{m[\underline{i}} \tilde\partial^{\underline{j}]} B_{mk}+ \beta^{im}\beta^{jn} H_{mnk}
}
and the non-geometric $R$-flux
\eq{
  R^{ijk}\defeq\, &3 \tilde\partial^{[\underline{i}} \beta^{\underline{jk}]}+
               3 \beta^{[\underline{i}m} {\partial}_m
                 \beta^{\underline{jk}]}+3 B_{mn} \beta^{[\underline{i}m} \tilde{\partial}^n \beta^{\underline{jk}]}
     +3\beta^{[\underline{i}m} \beta^{\underline{j}n}
       \tilde\partial^{\underline{k}]} B_{mn} \\
   &+\beta^{im}\beta^{jn}\beta^{kp}  H_{mnp} \, .
}

The Jacobi-identities for the brackets \eqref{commalgebra}
are trivial identities for the fluxes and can therefore be
considered as  their Bianchi identities. Again employing the
strong constraint, one arrives at the five independent relations
\eq{
\label{bianchids}
   {\cal D}_{[\underline{a}} H_{\underline{bcd}]}-{\textstyle {3\over 2}}
     H_{m[\underline{ab}} F^{m}{}_{\underline{cd}]}&=0  \\
   -{\textstyle {1\over 3}}\tilde{\cal D}^d H_{abc} + {\cal D}_{[\underline{a}} F^{d}{}_{\underline{bc}]}+
      F^{m}{}_{[\underline{bc}}  F^{d}{}_{\underline{a}]m}+ H_{m[\underline{ab}} Q_{\underline{c}]}{}^{md}&=0\\
 2 \tilde{\cal D}^{[\underline{c}} F^{\underline{d}]}{}_{[\underline{ab}]}+
      2 {\cal D}_{[\underline{a}} Q_{\underline{b}]}{}^{[\underline{cd}]} -
      F^{m}{}_{[\underline{ab}]} Q_{m}{}^{[\underline{cd}]} +
     4\, F^{[\underline{c}}{}_{m[\underline{a}} Q_{\underline{b}]}{}^{\underline{d}]m}-H_{abm} R^{mcd}  &=0\\
 -{\textstyle {1\over 3}}{\cal D}_d R_{abc} + \tilde{\cal D}^{[\underline{a}} Q_{d}{}^{\underline{bc}]}+
      Q_{m}{}^{[\underline{bc}}  Q_{d}{}^{\underline{a}]m}+ R^{m[\underline{ab}}
  F^{\underline{c}]}{}_{md}&=0\\
  \tilde{\cal D}^{[\underline{a}} R^{\underline{bcd}]}-{\textstyle {3\over 2}}
     R^{m[\underline{ab}} Q_{m}{}^{\underline{cd}]}&=0  .
}
These relations constitute the generalization of the fluxes and
their Bianchi identities to DFT. 

\subsection{Coexistence of fluxes for orientifolds}

In the previous derivation we have solely employed the strong constraint but
did not use any further constraint on  the $B$-field and the $\beta$-field.
Clearly, in a given patch string theory requires that 
only half of the degrees of freedom of these fields  field can be
independent. 
In addition, in the ten-dimensional theory one would require the
strong constraint to be satisfied.

First we notice that,  for tadpole cancellation, we need to perform an
orientifold projection so that an interesting  question to ask is
which orientifold even components of the
four types  of fluxes can be turned on simultaneously. 
The  details also depend on the concrete
orientifold projection. Thus, let us first consider
the behavior of the fields and fluxes under the
world-sheet parity transformation $\Omega: (\sigma,\tau)\to (-\sigma,\tau)$.
Here, the metric is invariant whereas  the two-form $B$ and the
bi-vector $\beta$ are anti-invariant. Moreover, since $\Omega$ maps
the winding number to its inverse, this should also hold
for the winding coordinate and the corresponding partial
derivative. Thus, we have
\eq{
     \Omega:\begin{cases} \partial_a\to \partial_a\, , \quad
                     \tilde \partial^a\to - \tilde \partial^a\\
                     B_{ab}\to -B_{ab} \, ,\quad
                            \beta^{ab}\to -\beta^{ab}\\
                               f^a{}_{bc}\to f^a{}_{bc}\,, \quad
                              \tilde f_a{}^{bc}\to -\tilde f_a{}^{bc}
              \end{cases}
}
so that the fluxes transform as
\eq{
\Omega:\begin{cases}     H_{abc}\to -H_{abc} \\
                         F^a{}_{bc}\to F^a{}_{bc} \\
                            Q_a{}^{bc}\to -Q_a{}^{bc}\\
                  R^{abc}\to R^{abc}\; .
              \end{cases}
}
Thus, under $\Omega$ only the fluxes $F$ and $R$ are even.
Dressing $\Omega$ by some $\mathbb Z_2$ space-time symmetries
introduces some extra minus signs along certain (reflected) legs,
but is not expected to alter the general structure of the number of
orientifold even fluxes to be turned on simultaneously.

As it will not affect the computation performed in the next section,
a complete  analysis about which fluxes can be simultaneously turned
on in a true string theory vacuum is beyond the scope of this paper. This question   has been under debate 
recently, where for instance it has been pointed out 
that by a Scherk-Schwarz reduction of DFT 
\cite{Aldazabal:2011nj,Geissbuhler:2011mx,Grana:2012rr,Dibitetto:2012rk}
all fluxes can appear,
but at the expense of weakening the strong constraint \eqref{strong_c}.
Let us 
emphasize again that the analysis in the next section goes 
through without imposing any  extra constraint on the fluxes
beyond their Bianchi identities.

\section{Dimensional oxidation}
\label{sec_three}

In this section we will start with the type IIA four-dimensional
superpotential \eqref{QRsuperpot} and analyze its ten-dimensional origin.
Concretely, as in \cite{Villadoro:2005cu}
we will consider a toroidal orientifold
for which we turn on all orientifold even  geometric and non-geometric NS-NS
fluxes, as well as all R-R fluxes.  
Computing the resulting four-dimensional scalar potential via
{\it Mathematica}, in a step by step procedure we will inspect  the
underlying ten-dimensional terms, whose dimensional reduction 
leads to precisely the terms present in that scalar potential.
This tedious analysis  gets complicated by the presence
of the Bianchi identities and tadpole conditions, which
have to be invoked heavily during the course of the computation.   

Let us emphasize that in this approach one has to clearly distinguish
objects with background values (like vacuum expectation values for
the fluxes) from objects that are moduli of the unfluxed Calabi-Yau
compactification (like complex structure and K\"ahler moduli). 
We will treat the former as generic constant parameters that
are only constrained by  the Bianchi identities. 
Thus, we are not imposing any of the possible DFT constraints discussed
in the previous section. We allow all of them to be present
at the same time.
Contrarily, for the initial compactification on the torus, 
we are working
in the $\tilde\partial_i=0$ DFT frame with the generalized metric
parametrized by $(G,B)$. As we will see, this mixed approach allows
us to detect non-trivial terms in the ten-dimensional action
that go beyond the usual supergravity action
\eqref{actiontypea}.

\subsection{A type IIA orientifold}

As our example we choose the  familiar type IIA
$\Omega I_3 (-1)^{F_L}$
orientifold on the orbifold $T^6/\mathbb Z_2\times \mathbb Z_2$.
Introducing on $T^6$ the coordinates
\eq{
    z^1=R^1 x^1+i\, R^2 x^2\, ,\qquad z^2=R^3 x^3+i\, R^4 x^4\, ,\qquad
    z^3=R^5 x^5+i\, R^6 x^6\, ,
}
where $0\le x^i\le 1$ and $R^i$ denote the circumference of the $i$-th circle.
The two $\mathbb Z_2$ actions are
\eq{ 
\label{thetaactions}
\theta:(z^1,z^2,z^3)\to (-z^1,-z^2,z^3)\\
     \ov\theta:(z^1,z^2,z^3)\to (z^1,-z^2,-z^3)\, ,
}
while $I_3$ acts as
\eq{
           I_3: (z^1,z^2,z^3)\to (-\ov{z}^1,-\ov{z}^2,-\ov{z}^3)\, .
}
The Hodge numbers of the  $T^6/\mathbb Z_2\times \mathbb Z_2$
orbifold are $(h^{21},h^{11})=(3,51)$,
but here we are only considering the untwisted sector with
$(h_{\rm ut}^{21},h_{\rm ut}^{11})=(3,3)$.
Due to the two $\mathbb Z_2$ symmetries,
the $T^6$ splits into a product of three $T^2$ tori, i.e.
in the untwisted sector of the orbifold we get three
complex structure and three K\"ahler moduli, i.e. one pair for each
$T^2$ factor
\eq{
         \hat u_1&={R^1/R^2}\,, \qquad \hat u_2={R^3/ R^4}\, ,\qquad
        \hat u_3={R^5/R^6}\\
       t_1&=R^1 R^2\, ,\qquad\hspace{0.3cm} t_2=R^3 R^4\, ,\qquad\hspace{0.3cm}  t_3=R^5 R^6\,   .
}
Let us choose the following basis of closed three-forms
\eq{
\label{formbasis}
       \alpha_0&=dx^1\wedge dx^3\wedge dx^5\,, \qquad\qquad
       \beta^0=dx^2\wedge dx^4\wedge dx^6\, ,\\
       \alpha_1&=dx^1\wedge dx^4\wedge dx^6\, , \qquad\qquad
       \beta^1=dx^2\wedge dx^3\wedge dx^5\, ,\\
       \alpha_2&=dx^2\wedge dx^3\wedge dx^6\, , \qquad\qquad
       \beta^2=dx^1\wedge dx^4\wedge dx^5\, ,\\
      \alpha_3&=dx^2\wedge dx^4\wedge dx^5\, ,\qquad\qquad
       \beta^3=dx^1\wedge dx^3\wedge dx^6\, 
}
satisfying $\int \alpha_I\wedge \beta^J=-\delta_I{}^J$. The
holomorphic three-form is $\Omega_3=dz^1\wedge dz^2\wedge dz^3$ so
that we can expand
\eq{
\Omega^c&={\rm Re} (i\, e^{-\phi}\, \Omega_3) - i C^{(3)} \\
  &=S\, \beta^0 - U_1\, \beta^1 - U_2\, \beta^2 - U_3\, \beta^3\, ,
}
where
\eq{\label{chiraltwoa}
            S&=e^{-\phi} R^2 R^4 R^6 -i C^{(3)}_{246}\, ,\qquad
    U_1=e^{-\phi}  R^2 R^3 R^5 +iC^{(3)}_{235}, \\
   U_2&= e^{-\phi}  R^1 R^4 R^5 +iC^{(3)}_{145}, \qquad U_3= e^{-\phi} R^1 R^3 R^6+iC^{(3)}_{136}
}
are the bosonic components of the chiral superfields for the
orientifold even complex structure moduli.
The chiral superfields for the complexified K\"ahler moduli are
\eq{
        T_1=t_1+i B_{12}, \qquad T_2=t_2+i B_{34}, \qquad T_3=t_3+i B_{56}\, .
 }
Then, the non-vanishing components of the internal  ten-dimensional
metric in string frame are
\eq{
     g_{MN}={\rm blockdiag}\Big({e^{2\phi}\over t_1 t_2 t_3} \tilde g_{\mu\nu},(R^1)^2,(R^2)^2,(R^3)^2,(R^4)^2,(R^5)^2,(R^6)^2\Big)
}
where here $\tilde g_{\mu\nu}$ denotes the four-dimensional metric in
Einstein-frame. The internal components 
can be expressed as
\eq{
g_{11}&=t_1 \sqrt{\frac{u_2 u_3}{s u_1}}\, , \qquad g_{22}=t_1 \sqrt{\frac{s
    u_1}{u_2 u_3}}\, , \qquad
g_{33}=t_2 \sqrt{\frac{u_1 u_3}{s u_2}}\, ,\\
g_{44}&=t_2 \sqrt{\frac{s u_2}{u_1 u_3}}\, , \qquad g_{55}=t_3 \sqrt{\frac{u_2
    u_1}{s u_3}}\, , \qquad g_{66}=t_3 \sqrt{\frac{s u_3}{u_2 u_1}}\, ,
}
where $s$ and the $u_i$ are the real components of the complex
fields in \eqref{chiraltwoa}.
The tree-level K\"ahler potential for the seven moduli fields is given by
\eq{
K = -\ln\left( \frac{S +\ov{S}}{2}\right) -\sum_{i=1,2,3}\ln\left( \frac{U_i
  +\ov{U_i}}{2}\right)- \, \sum_{i=1,2,3}\ln\left( \frac{T_i
  +\ov{T_i}}{2}\right) \, .
}
The independent, orientifold even components of the NS-NS background fluxes are
\eq{
\label{fluxindep}
 \ov{H}_{ijk}:\qquad &\ov{H}_{135}\,,\  \ov{H}_{146}\, ,\  \ov{H}_{236}\, ,\  \ov{H}_{245}\\[0.1cm]
    \ov{F}^k{}_{ij}:\qquad &\ov{F}^6{}_{13}\,,\   \ov{F}^5{}_{23}\,,\   \ov{F}^6{}_{24}\,,\   \ov{F}^5{}_{14}\\
            &\ov{F}^2{}_{35}\,,\   \ov{F}^1{}_{45}\,,\   \ov{F}^2{}_{46}\,,\   \ov{F}^1{}_{36}\\
            &\ov{F}^4{}_{51}\,,\   \ov{F}^3{}_{61}\,,\   \ov{F}^4{}_{62}\,,\   \ov{F}^3{}_{52}\\[0.1cm]
    \ov{Q}_k{}^{ij}:\qquad &\ov{Q}_1{}^{35}\,,\   \ov{Q}_2{}^{45}\,,\   \ov{Q}_1{}^{46}\,,\   \ov{Q}_2{}^{36}\\
            &\ov{Q}_5{}^{13}\,,\   \ov{Q}_6{}^{23}\,,\   \ov{Q}_5{}^{24}\,,\   \ov{Q}_6{}^{14}\\
            &\ov{Q}_3{}^{51}\,,\   \ov{Q}_4{}^{61}\,,\   \ov{Q}_3{}^{62}\,,\   \ov{Q}_4{}^{52}\\[0.1cm]
   \ov{R}^{ijk}:\qquad  &\ov{R}^{246}\,,\  \ov{R}^{235}\, ,\  \ov{R}^{145}\,
   ,\  \ov{R}^{136}\, .
}
In the R-R sector the orientifold even fluxes are
\eq{  \overline{G}^{(0)},\; \overline{G}^{(2)}_{12},\;  \overline{G}^{(2)}_{34},\;
   \overline{G}^{(2)}_{56},\; \overline{G}^{(4)}_{1234},\;
   \overline{G}^{(4)}_{1256},\; \overline{G}^{(4)}_{3456},\;
   \overline{G}^{(6)}_{123456}\, .
}
The detailed  form of the superpotential
\eq{
         W_{\rm NS}=-{i\over 4} \int_X \overline{\mathfrak{H}}^C\wedge \Omega^c
   +{1\over 4}\, \int_{X} e^{iJ_c} \wedge \ov{G} \,\\[0.1cm]
}
in terms of the fluxes and moduli can be found in appendix \ref{appendix_0}

\subsection{Oxidation to 10D  type IIA action}

Taking now the superpotential \eqref{superpotex}
and computing the  scalar F-term potential, 
the procedure described in the beginning of section \ref{sec_three} 
reveals that the result  can be obtained from a couple of
generalized kinetic terms in a ten-dimensional action
\eq{
\label{oxiaction}
        S={1\over 2}\int  d^{10} x\,  \sqrt{-g} \Big(  {\cal L}^{\rm NS}_1
           +{\cal L}^{\rm NS}_2 +{\cal L}^{\rm R} \Big)\, .
}
In order to express the result it turns out to be convenient
to introduce the following combinations or orbits of fluxes
\eq{
\label{orbitfluxes}
     \mathfrak{H}_{ijk}&=\ov{H}_{ijk}+3\, \ov{F}^m{}_{[\underline{ij}}\,
       B_{m\underline{k}]}
         +3\, \ov{Q}_{[\underline{i}}{}^{mn} B_{m\underline{j}}\,
           B_{n\underline{k}]}
    +\ov{R}^{mnp}  B_{m[\underline{i}} B_{n\underline{j}} B_{p\underline{k}]}\\
      \mathfrak{F}^i{}_{jk}&=\ov{F}^i{}_{jk}+2\,\ov{Q}_{[\underline{j}}{}^{mi}  B_{m\underline{k}]}\,
    +\ov{R}^{mni}  B_{m[\underline{j}} B_{n\underline{k}]}\\
    \mathfrak{Q}_k{}^{ij}&=\ov{Q}_k{}^{ij}+\ov{R}^{mij}  B_{mk}\\
     \mathfrak{R}^{ijk}&=\ov{R}^{ijk}\, .
       }
Here by the overlining 
we indicated which fields in the dimensional reduction are 
treated as backgrounds and which as moduli.
Then, we oxidize a term containing three metric factors
\eq{
     {\cal L}^{\rm NS}_1=-{e^{-2\phi}\over 12}\Big(  & {\mathfrak{H}}_{ijk}\,  {\mathfrak{H}}_{i'j'k'}\, g^{ii'}
                           g^{jj'} g^{kk'} +
                         3\, {\mathfrak{F}}^i{}_{jk}\, {\mathfrak{F}}^{i'}{}_{j'k'}\,
                         g_{ii'} g^{jj'} g^{kk'}\\
                     &+3\, {\mathfrak{Q}}_k{}^{ij}\, {\mathfrak{Q}}_{k'}{}^{i'j'}\,
                         g_{ii'} g_{jj'} g^{kk'}
                       +{\mathfrak{R}}^{ijk} \, {\mathfrak{R}}^{i'j'k'}\, g_{ii'}
                           g_{jj'} g_{kk'} \Big)
}
and a term  containing  a single metric factor
\eq{
    {\cal L}^{\rm NS}_2=-{e^{-2\phi}\over 2}\Big( & {\mathfrak{F}}^m{}_{ni}\, {\mathfrak{F}}^{n}{}_{mi'}\,
                           g^{ii'} +
                      {\mathfrak{Q}}_m{}^{ni}\, {\mathfrak{Q}}_{n}{}^{mi'}\,
                           g_{ii'} \\
                       &-  {\mathfrak{H}}_{mni} \, {\mathfrak{Q}}_{i'}{}^{mn}\,
                      g^{ii'} - {\mathfrak{F}}^i{}_{mn}\, {\mathfrak{R}}^{mni'} \,
                       g_{ii'}\Big)\, .
}
The contribution from the Ramond-Ramond-sector is
\eq{
 {\cal L}^{\rm R}=  -\frac{1}{2} \sum_{p = 0,2,4,6} |G^{(p)}|^2\, ,
}
where the $p$-form field strengths are defined as
\eq{
      G^{(p)}={1\over p!} G^{(p)}_{i_1\ldots i_p}\, dx^{i_1}\wedge \ldots \wedge
      dx^{i_p}}
with the components
\eq{
\label{RRoxidized}
   G^{(0)}&=\overline{G}^{(0)} +{1\over 6}\, {\mathfrak R}^{mnp} C^{(3)}_{mnp}\\[0.1cm]
   G^{(2)}_{ij}&=\overline{G}^{(2)}_{ij} - B_{ij} \overline{G}^{(0)}
                +{\mathfrak Q}_{[\underline{i}}{}^{mn}
                  C^{(3)}_{mn\underline{j}]}\\[0.1cm]
  G^{(4)}_{ijkl}&=\overline{G}^{(4)}_{ijkl} - 6\, B_{[\underline{ij}}
    \overline{G}^{(2)}_{\underline{kl}]}+3\,B_{[\underline{ij}}
    B_{\underline{kl}]}\, \overline{G}^{(0)}-
   6\, {\mathfrak F}^m{}_{[\underline{ij}}
                  C^{(3)}_{m\underline{kl}]}\\[0.1cm]
  G^{(6)}_{ijklmn}&=\overline{G}^{(6)}_{ijklmn} - 15\, B_{[\underline{ij}}
    \overline{G}^{(4)}_{\underline{klmn}]}+45\, B_{[\underline{ij}}
    B_{\underline{kl}}\, \overline{G}^{(2)}_{\underline{mn}]}\\
   &\phantom{aaaaaaaaaaaaaaaaaaa}
  -15\, B_{[\underline{ij}}
    B_{\underline{kl}} B_{\underline{mn}]}\, \overline{G}^{(0)}
   -20\, {\mathfrak H}_{[\underline{ijk}}
                  C^{(3)}_{\underline{lmn}]}\,  .
}
Here we included the background flux $G=e^{-B}\ov{G}$.
Taking now the action \eqref{oxiaction} and dimensionally reducing
it to four-dimensions gives a scalar potential which is a
sum of the desired F-term and a D-term. Note that 
the F-term scalar potential has several additional terms, 
which upon invoking the
Bianchi identities for constant fluxes
\eq{
\label{bianchids1}
        \ov{H}_{m[\underline{ab}} \ov{F}^{m}{}_{\underline{cd}]}&=0  \\
         \ov{F}^{m}{}_{[\underline{bc}}  \, \ov{F}^{d}{}_{\underline{a}]m}+ \ov{H}_{m[\underline{ab}} \, \ov{Q}_{\underline{c}]}{}^{md}&=0\\
      \ov{F}^{m}{}_{[\underline{ab}]} \, \ov{Q}_{m}{}^{[\underline{cd}]} -
     4\, \ov{F}^{[\underline{c}}{}_{m[\underline{a}} \, \ov{Q}_{\underline{b}]}{}^{\underline{d}]m} + \ov{H}_{mab} \, \ov{R}^{mcd}  &=0\\
       \ov{Q}_{m}{}^{[\underline{bc}}  \, \ov{Q}_{d}{}^{\underline{a}]m}+ \ov{R}^{m[\underline{ab}} \, \,
  \ov{F}^{\underline{c}]}{}_{md}&=0\\
      \ov{R}^{m[\underline{ab}} \, \ov{Q}_{m}{}^{\underline{cd}]}&=0  .
}
are nullified. More details on this
computation are presented in appendix \ref{appendix_a}.

In order to describe the D-term let us define a three-form
\eq{
         \tau={1\over 6} \tau_{ijk} dx^i\wedge dx^j\wedge dx^k}
with
\eq{
  \tau_{ijk}=\ov{H}_{ijk} \, \overline{G}^{(0)} +3\, \ov{F}^m{}_{[\underline{ij}}
                  \overline{G}^{(2)}_{m\underline{k}]}-{3\over 2}
              \ov{Q}_{[\underline{i}}{}^{mn}
                  \overline{G}^{(4)}_{mn\underline{jk}]}-{1\over 6}
               \ov{R}^{mnp}  \overline{G}^{(6)}_{mnpijk}\, .
}
Then the D-term is
\eq{
   V_D = -\frac{1}{2}e^K t_1 t_2 t_3\Big[s\, \tau_{135} - u_1\, \tau_{146}  - u_2\, \tau_{236}  - u_3\, \tau_{245}\Big]
}
which is a contribution to the NS-NS tadpole. 
As in \cite{Villadoro:2005cu}, due to R-R tadpole cancellation
we expect this term to be canceled against the tensions of the D6-branes and
O6-planes. From this one can deduce the existence of a Chern-Simons term
\eq{
           S_{\rm CS}\sim \int C^{(7)}\wedge \tau
}
in the ten-dimensional action, which was also suggested in \cite{Aldazabal:2006up}.

\subsection{A type IIB orientifold}

As a second explicit example, we now perform the analogous
computation for a T-dual type IIB orientifold.
Applying a T-duality in the three-directions $x_2,x_4,x_6$, leads
to the type IIB  orientifold with action
$\Omega\,  I_6 \, (-1)^{F_L}$  on the orbifold
$ T^6 / \mathbb Z_2\times \mathbb Z_2$. 
The fixed point set  of this orientifold gives  $O3/O7$-planes.
The complex coordinates $z_i$'s on $T^6=T^2\times T^2\times T^2$ are defined as
\eq{
z^1=x^1+ i \, U_1 \,  x^2, \qquad z^2=x^3+ i \, U_2 \, x^4,\qquad z^3=x^5+ i
\, U_3 \, x^6\, ,
}
where the three complex structure moduli $U_i$'s can be written as 
$U_i= u_i + i\, v_i,\,\,i=1,2,3.$ 
The orbifold symmetry of the two $\mathbb Z_2$ acts as in \eqref{thetaactions}
while the involution $I_6$ is
\eq{
I_6 : (z^1,z^2,z^3)\rightarrow (-z^1,-z^2,-z^3)\, .
}
Choosing the basis of closed three-forms \eqref{formbasis},
the holomorphic 3-form $\Omega_3$ can be expanded as
\eq{
\Omega_3=  \alpha_0 + i \, (U_1 \beta^1 +&U_2 \beta^2 + U_3 \beta^3)-i U_1
U_2 U_3 \beta^0\\[0.1cm] 
& -U_2 U_3 \alpha_1- U_1 U_3 \alpha_2 - U_1 U_2 \alpha_3\, .
}
In addition one has the axio-dilaton chiral superfield, whose bosonic
component is
\eq{
 S=& \, e^{-\phi}-i \, C^{(0)}\, .
}
The chiral superfields related to the K\"ahler moduli
are generically encoded in  the complexified  four-cycle volumes
\eq{
J^c= \frac{1}{2} e^{-\phi} J \wedge J +i \, C^{(4)}\; .
}
In our case, these moduli are
 \eq{
T_1 = \, \tau_1+ i \, C^{(4)}_{3456} \, ,\qquad
T_2 = \, \tau_2+ i \, C^{(4)}_{1256} \, ,\qquad
T_3 = \, \tau_3+ i \, C^{(4)}_{1234} \, ,
 }
where the real parts can be expressed in terms of the two-cycle volumes $t_i$
as
$\tau_1 = e^{-\phi} \, t_2 \, t_3$, $\tau_2 = e^{-\phi} \, t_3 \, t_1$ and
$\tau_3 = e^{-\phi}  \, t_1 \, t_2$. 
We also need to express the two-cycle volumes $t_i$ in terms of the 
four-cycles volumes $\tau_i$
\eq{
\label{twoinfour}
    t_1=\sqrt{\tau_2\,  \tau_3\over \tau_1\, s}\,, \qquad
    t_2=\sqrt{\tau_1\,  \tau_3\over \tau_2\, s}\,, \qquad
   t_3=\sqrt{\tau_1\,  \tau_2\over \tau_3\, s}\, 
}
with $s={\rm Re}(S)$.
Now, the non-vanishing components of the metric
in string frame are
\eq{
     g_{MN}={\rm blockdiag}\Big({e^{\phi\over 2}\over {\sqrt{\tau_1\, \tau_2
           \, \tau_3}}} \, \, \tilde g_{\mu\nu}, \, \, g_{ij}\Big) \, .
}
Further, the string frame internal metric $g_{ij}$  
is also block-diagonal and 
has the following non-vanishing components,
\eq{
g_{11}&=\frac{t_1}{u_1}\,,\quad    g_{12}=-\frac{t_1 v_1}{u_1} =g_{21}\, , \quad
g_{22}=\frac{t_1(u_1^2+v_1^2)}{u_1}\, ,\\
g_{33}&=\frac{t_2}{u_2}\, ,\quad    g_{34}=-\frac{t_2 v_2}{u_2}=g_{43}\, , \quad
g_{44}=\frac{t_2(u_2^2+v_2^2)}{u_2}\, ,\\
g_{55}&=\frac{t_3}{u_3}\,, \quad    g_{56}=-\frac{t_3 v_3}{u_3}=g_{65}\,, \quad
g_{66}=\frac{t_3(u_3^2+v_3^2)}{u_3}\, .\\
}
Using \eqref{twoinfour} all components can also be expressed
in term of the non-axionic  components of the seven chiral
superfields $S,T_{1,2,3},U_{1,2,3}$.

Since the background fluxes $F^i{}_{jk}$ and $R^{ijk}$ are odd under
the orientifold projection, the only invariant fluxes
are the following components of the three-forms $H$ and $G^{(3)}$
\eq{
     \ov{H}:\quad &{\ov H}_{135}\,,\  {\ov H}_{146}\, ,\  {\ov H}_{236}\, ,\  {\ov H}_{245}\\
            &{\ov H}_{246}\,, \ {\ov H}_{235}\, ,\ {\ov H}_{145}\, ,\ {\ov
       H}_{136}\, ,\\[0.1cm]
    \ov G^{(3)}:\quad & {\ov G}_{135}\,,\  {\ov G}_{146}\, ,\  {\ov G}_{236}\, ,\  {\ov G}_{245}\\
            &{\ov G}_{246}\,, \ {\ov G}_{235}\, ,\ {\ov G}_{145}\, ,\ {\ov
      G}_{136}
}
and the  components of non-geometric $Q$-flux\footnote{For S-dual completion of type IIB orientifolds, one needs to introduce additional non-geometric P-fluxes which are R-R analogue of non-geometric Q-fluxes \cite{Aldazabal:2008zza}.}
\eq{
\ov Q:\quad &\ov Q_1{}^{35}\,,\   \ov Q_2{}^{45}\,,\   \ov
Q_1{}^{46}\,,\   \ov Q_2{}^{36}\\
            &\ov Q_5{}^{13}\,,\   \ov Q_6{}^{23}\,,\   \ov
Q_5{}^{24}\,,\   \ov Q_6{}^{14}\\
            &\ov Q_3{}^{51}\,,\   \ov Q_4{}^{61}\,,\   \ov
Q_3{}^{62}\,,\   \ov Q_4{}^{52}\\
            &\ov Q_2{}^{35}\,,\   \ov Q_5{}^{23}\,,\   \ov
Q_3{}^{52}\,,\   \ov Q_2{}^{46}\\
            &\ov Q_4{}^{51}\,,\   \ov Q_1{}^{45}\,,\   \ov
Q_5{}^{14}\,,\   \ov Q_4{}^{62}\\
            &\ov Q_6{}^{13}\,,\   \ov Q_3{}^{61}\,,\   \ov
Q_1{}^{36}\,,\   \ov Q_6{}^{24}\; .
}
The full superpotential is given by
\eq{
\label{typeIIBW}
W = -\frac{i}{4}\int_{X} \left( S \, \ov{H}+ \, \ov{Q}\cdot
J^c\right)\wedge \Omega_3 + \frac{1}{4} \int_X \ov G^{(3)} \wedge \Omega_3 = 
W_{NS} + W_{R}
}
where the three-form $\ov{Q}\cdot J^c={1\over 6} (\ov{Q}\cdot J^c)_{ijk}\,
dx^i\wedge dx^j\wedge dx^k$ is defined as
\eq{
    (\ov{Q}\cdot J^c)_{ijk}={3\over 2}\, \ov{Q}_{[\underline{i}}{}^{mn\,}
    J^c_{mn\underline{jk}]}\, .
}
The explicit form of this superpotential is presented in appendix \ref{appendix_0}.
Together with the   K\"ahler potential 
\eq{
\label{typeIIBK}
K &= -\ln\left( \frac{S +\ov{S}}{2}\right) -\sum_{i=1,2,3}\ln\left( \frac{U_i +\ov{U_i}}{2}\right)- \, \sum_{i=1,2,3}\ln\left( \frac{T_i +\ov{T_i}}{2}\right) 
}
it allows now to compute the effective four-dimensional scalar potential.

\subsection{Oxidation to 10D  type IIB action}

Similar to the previous type IIA case, a close inspection
of the resulting scalar potential reveals that it can be obtained 
via dimensional reduction from a couple of
generalized kinetic terms in a ten-dimensional action
\eq{
\label{oxiaction2}
        S={1\over 2}\int  d^{10} x\,  \sqrt{-g} \Big(  {\cal L}^{\rm NS}_1
           +{\cal L}^{\rm NS}_2 +{\cal L}^{\rm R} \Big)}
with the oxidized  terms  given as 
\eq{
\label{tyepIIBoxidize}
{\cal L}^{\rm NS}_1&=-{e^{-2\phi}\over 12}\Big( \ov{H}_{ijk}\,  \ov{H}_{i'j'k'}\, g^{ii'}
                           g^{jj'} g^{kk'} +3\, \ov{Q}_k{}^{ij}\, \ov{Q}_{k'}{}^{i'j'}\,g_{ii'} g_{jj'} g^{kk'} \Big) \\
  {\cal L}^{\rm NS}_2&=-{e^{-2\phi}\over 2}\Big(  \ov{Q}_m{}^{ni}\, \ov{Q}_{n}{}^{mi'}\, g_{ii'} -  \ov{H}_{mni} \, \ov{Q}_{i'}{}^{mn}\, g^{ii'} \Big) \\
 {\cal L}^{\rm R}&=  -{1\over 12}\, G^{(3)}_{ijk}\,  G^{(3)}_{i'j'k'}\, g^{ii'}
  g^{jj'} g^{kk'} 
}
with
\eq{
\label{g3formel}
G^{(3)}_{ijk}=\ov{G}_{ijk}- C^{(0)} \, \ov{H}_{ijk} +\frac{3}{2}\,
\ov{Q}_{[\underline{i}}{}^{lm} C^{(4)}_{lm\underline{jk}]}\, .
}
Let us mention that for our example there are no two-forms
anti-invariant under the orientifold projection
so that  no $B_2$ and $C_2$ moduli are present. 
As a consequence, the type IIA flux orbits like  $\mathfrak{H}_{ijk}$ and 
$\mathfrak{Q}_k{}^{ij}$ simplify to $H_{ijk}$ and $Q_k{}^{ij}$.
The redefined three-form \eqref{g3formel} 
also appeared in \cite{Font:2008vd}.

As for the type IIA case, invoking the Bianchi identities for
constant fluxes \eqref{bianchids1}, the dimensional reduction of the above kinetic
terms leads to the F-term scalar potential induced by the superpotential
and some additional D-terms. More details on this lengthy computation
can be found in appendix \ref{appendix_a}.
There arises a D-term 
\eq{
V_{D3} = \frac{1}{2} \, e^K \, u_1 u_2 u_3 \, \Big[ 20\, \ov{G}_{[\underline{123}}
                  \ov H_{\underline{456}]} \,s \Big]
}
corresponding to the D3-brane tadpole and a D-term 
\eq{
V_{D7} = -\frac{1}{2} \, e^K \, u_1 u_2 u_3\, \, 
\Big[\ov Q_{[\underline{1}}{}^{jk} \, \ov G_{jk\underline{2}]} \, \tau_1+ \ov
Q_{[\underline{3}}{}^{jk} \, \ov G_{jk\underline{4}]}\, \tau_2+ \ov
Q_{[\underline{5}}{}^{jk} \, \ov G_{jk\underline{6}]} \, \tau_3 \Big]
}
corresponding to the three D7-brane tadpoles.
Since these are related to the R-R tadpoles, there should also exist
CS-terms  
\eq{ 
S_{CS}\sim -\int C^{(4)} \wedge G \wedge H +\int C^{(8)} \wedge Q\cdot
G\, ,
}
where the first one is familiar from the standard type IIB supergravity action.

\section{Relation to DFT}

Let us recall that for both the type IIA and the type IIB orientifold
we have succeeded to extract  ten-dimensional actions \eqref{oxiaction} and 
\eqref{oxiaction2}, whose
dimensional reduction induces precisely  the F-term scalar potential
implied by the flux induced superpotential and additional D-term
potentials, which are related to the tadpole cancellation conditions.
Thus, the program of the derivation of the Gukov-Vafa-Witten superpotential
first performed  in \cite{Taylor:1999ii} 
could  be realized also for  more generic fluxes. The question now
is whether the ten-dimensional actions are  really related to the main
candidate for an action including both geometric and non-geometric fluxes,
i.e. the DFT action \eqref{dftaction}.  
In this section, let us discuss this in more detail,
where we consider the NS-NS and the R-R part separately.

\subsection{Relating the oxidized NS-NS action to DFT}

As in the discussions of Scherk-Schwarz reductions of DFT in 
\cite{Aldazabal:2011nj,Geissbuhler:2011mx,Grana:2012rr,Geissbuhler:2013uka,Berman:2012uy,Berman:2013cli},
it is convenient to  introduce the DFT fluxes ${\cal F}_{IJK}$ with
\eq{
     {\cal F}_{ijk}=H_{ijk}\, , \quad {\cal F}^i{}_{jk}=F^i{}_{jk}\, , \quad 
   {\cal F}_k{}^{ij}=Q_k{}^{ij}\, , \quad  {\cal F}^{ijk}=R^{ijk}\, .
}
Then, we observe that the NS-NS sector action can be compactly written as 
\eq{
 \label{dftactionflux}
{\cal L}^{\rm NS}_1+{\cal L}^{\rm NS}_2=e^{-2\phi}
 {\cal F}_{IJK} {\cal F}_{I'J'K'}\,
\Big( \frac{1}{4} {\cal H}^{II'} \eta^{JJ'} \eta^{KK'} 
 -\frac{1}{12} {\cal H}^{II'} {\cal H}^{JJ'} {\cal H}^{KK'}\Big)\, .
}
Note that the off-diagonal components of the generalized metric ${\cal H}$
automatically generate those flux orbits \eqref{orbitfluxes}
$\mathfrak{H}_{ijk}$, $\mathfrak{F}^i{}_{jk}$, $\mathfrak{Q}_k{}^{ij}$
and $\mathfrak{R}^{ijk}$. This form of the action is very reminiscent of the 
flux formulation  of the DFT action, which in a flat frame can be expressed
as follows
\eq{
 \label{dftactionfluxb}
   S_{{\rm DFT}}={1\over 2}
\int
d^{20}X \; &e^{-2d}\, \bigg[ {\cal F}_{ABC} {\cal F}_{A'B'C'}\,
\Big( \frac{1}{4} {S}^{AA'} \eta^{BB'} \eta^{CC'} \\
&  -\frac{1}{12} {S}^{AA'} {S}^{BB'} {S}^{CC'}
    -\frac{1}{6} {\eta}^{AA'} \eta^{BB'} \eta^{CC'} \Big)\\
& + {\cal F}_A {\cal F}_{A'}\Big( \eta^{AA'}-{S}^{AA'} \Big)\bigg]\, .
}
Here ${\cal F}_A$ is defined as
\eq{
     {\cal F}_A=\Omega^B{}_{BA}+2 E_A{}^I \partial_I d
}
with the generalized Weitzenb\"ock connection
\eq{
      \Omega_{ABC}=E_A{}^I \partial_I E_B{}^J\, E_{CJ}\, .
}
In this form the action is motivated by the  scalar
potential appearing in gauged supergravity.
In performing the dimensional reduction of \eqref{dftactionfluxb}, we allow
generic constant background fluxes $\ov{\cal F}_{ABC}$. Moreover,   
for the dynamical fields we solve
the strong constraint via $\tilde\partial^i=0$ and 
use the generalized non-holonomic frame \eqref{gennonhol}.
Then, one can compute for the components of ${\cal F}_{ABC}$
\eq{
\label{fluxflatcurved}
     {\cal F}_{abc}&=e_a{}^i\,e_b{}^j\,e_c{}^k\; {\mathfrak H}_{ijk}\, ,\qquad
    {\cal F}^a{}_{bc}=e^a{}_i\,e_b{}^j\,e_c{}^k\; {\mathfrak F}^i{}_{jk}\, ,\\
    {\cal F}_c{}^{ab}&=e^a{}_i\,e^b{}_j\,e_c{}^k\; {\mathfrak Q}_k{}^{ij}\, ,\qquad
    {\cal F}^{abc}=e^a{}_i\,e^b{}_j\,e^c{}_k\; {\mathfrak R}^{ijk}\; 
} 
showing that the flux orbits \eqref{orbitfluxes} arise automatically from
the off-diagonal component of the generalized  non-holonomic frame
\eqref{gennonhol}.

Thus, for type IIA the first two terms in \eqref{dftactionfluxb} precisely
match our oxidized NS-NS action \eqref{dftactionflux}.
The additional three terms in \eqref{dftactionfluxb} cannot be
detected by our computation.  On the one hand, the 
term  ${\cal F}_{ABC} {\cal F}^{ABC}$ vanishes for orientifold even fluxes
and, on the other hand, ${\cal F}_A$ vanishes as we 
only have fluxes with precisely one leg on 
each $T^2$ as well as  constant metric and dilaton.
We conclude that the oxidized type IIA NS-NS sector action 
is indeed compatible
with the ten-dimensional DFT action in the NS-NS sector. 
The same holds for the type IIB
orientifold, where the result is even simpler as all internal 
components of the B-field  are modded out by the orientifold projection.

\subsection{Relating the oxidized R-R action to DFT}

Let us now investigate whether this continues to the R-R sector.
In order to compactly write the oxidized action, as shown in
\cite{Hohm:2011zr,Hohm:2011dv,Hohm:2011cp,Jeon:2012kd}, 
it is convenient to put the R-R fields into  the spinor
representation of $O(10,10)$. One defines the generalized  $\Gamma$-matrices
as $\Gamma^A=(\Gamma_a,\Gamma^a)$ where we remind the reader
that  $a$ is a flat index. 
Then all $\Gamma^A$ can be chosen to be real and
the Clifford algebra reads
\eq{
     \{\Gamma^A,\Gamma^B\}=\eta^{AB}\, .
}
Therefore,  the only non-vanishing anti-commutator is
$\{\Gamma^a,\Gamma_b\}=\delta^a{}_b$. Thus $\Gamma_a$ can
be considered as a fermionic lowering operator and $\Gamma^a$ as
the corresponding raising operator. Introducing a vacuum state $|0\rangle$
with $\Gamma_a|0\rangle=0$, we can put all R-R fields
in the spinor representation as
\eq{
{\cal G}=\sum_{n} {1\over n!}\, G^{(n)}_{i_1\ldots i_n}\, e_{a_1}{}^{i_1}\ldots 
        e_{a_n}{}^{i_n}\, \Gamma^{a_1\ldots a_n}|0\rangle\, ,
}
where as usual $\Gamma^{a_1\ldots a_n}$ defines the totally anti-symmetrized
product of $n$ $\Gamma$-matrices.
Similarly, we combine all the R-R gauge potentials $C^{(n)}$ into
a spinor ${\cal C}$.
Then, as shown in \cite{Geissbuhler:2011mx}, one 
can compactly define  the R-R field strengths as
\eq{
    {\cal G}=\slnabla {\cal C}
}
with the generalized fluxed Dirac operator defined as
\eq{
\label{newdirac}
    \slnabla=\Gamma^A D_A -{1\over 3} \Gamma^A {\cal F}_A-{1\over 6}
    \Gamma^{ABC} {\cal F}_{ABC}\; .
}
Evaluating this for the $G^{(3)}$-flux in the type IIB example, we find
\eq{
\label{hannover96}
   {1\over 3!} &G^{(3)}_{ijk}\, e_a{}^i\, e_b{}^j\, e_c{}^k\,
   \Gamma^{abc}|0\rangle =\Gamma^A D_A  \,{\cal C}-
   {1\over 6} \Gamma^{ABC} {\cal F}_{ABC} \,{\cal C}\\
   &={1 \over 2}\,\Gamma^a \partial_a  \,C^{(2)}_{bc} \Gamma^{bc}|0\rangle-{1\over 6} H_{abc} C^{(0)} \Gamma^{abc}|0\rangle\\
   &\hspace{1cm}-{1\over 6}\, {3\over 4!} \, \Gamma^m\Gamma_n\Gamma_p Q_m{}^{np}
   C^{(4)}_{ijkl}\, e_a{}^i\, e_b{}^j\, e_c{}^k \, e_d{}^l\,
   \Gamma^{abcd}|0\rangle\\
   &={1 \over 2}\, \partial_{i} C^{(2)}_{jk}\, e_a{}^i\, e_b{}^j\, e_c{}^k\,
   \Gamma^{abc}|0\rangle-{1\over 6}\, H_{ijk}\, C^{(0)} e_a{}^i\, e_b{}^j\,
   e_c{}^k\, \Gamma^{abc}|0\rangle\\
   &\hspace{1cm}+{1\over 6}\, {3\over 2} Q_a{}^{mn}
    C^{(4)}_{ijkl}\, e_b{}^i\, e_c{}^j\, e_m{}^k \, e_n{}^l\,
   \Gamma^{abc}|0\rangle\,,
}
from which one can conclude\footnote{In \eqref{hannover96} we were using that
  $f^a{}_{bc}=0$, as it is odd under orientifold projection.}
\eq{
G^{(3)}_{ijk}=\ov{G}_{ijk} -C^{(0)} \, \ov{H}_{ijk} +\frac{3}{2}\,
\ov{Q}_{[\underline{i}}{}^{lm} C^{(4)}_{lm\underline{jk}]}\, .
}      
This is precisely the oxidized relation \eqref{g3formel}.
Using \eqref{fluxflatcurved}, the last term in \eqref{newdirac}
also reproduces precisely the type IIA
NS-NS flux contributions in \eqref{RRoxidized}.
Therefore, we conclude that  the oxidized R-R action is compatible with
the proposed DFT action, as well.
We find it quite remarkable to realize how much information about the underlying
ten-dimensional action  is already contained in this innocently  
looking superpotential.

\section{Conclusions}

In this paper we have analyzed a couple of  aspects of
non-geometric flux compac\-tifications of the type II string
theories to four-dimensions in the light of recent progress
in DFT. 
First, we have rederived  the form of the four type
of fluxes $H,F,Q$ and $R$ in terms of the fundamental
fields appearing in the generalized metric of DFT.
Since the latter are subject to a number of constraints,
it appears to be a non-trivial question, which fluxes
one can truly turn on in a string theory vacuum. This complication
arises, as DFT in ten-dimensions contains too many degrees
of freedom so that it has to be equipped with some version of a 
strong constraint.

In the main part of the paper, 
we have explicitly verified that the type IIA and type IIB
superpotentials for constant geometric and non-geometric fluxes are
compatible with the recently proposed form of the tree-level DFT action
in both the NS-NS and the R-R sector. We would like to emphasize
that this result holds
independent of the actual realization of the non-vanishing fluxes, i.e.
prior to invoking any  additional constraints beyond the strong constraint,
which was employed for deriving the Bianchi identities.
We were investigating two concrete toroidal orientifolds, for which we
turned on constant orientifold even fluxes and then oxidized the resulting
F-term scalar potential to ten-dimensions. This led to a number of
kinetic terms for the fluxes, which showed precisely the intricate structure
present in the flux  DFT action, which was motivated by gauged supergravity. 
In this computation, the fluctuations of the massless
fields $G$ and $B$ were still treated in the usual geometric frame, whereas for
the backgrounds generic, i.e. also non-geometric, fluxes were allowed.
This found precise match provides further compelling
evidence  for both the correctness of the superpotential and
the DFT action. 

Due to our still restricted ansatz, it was not possible to see all DFT terms.
To also make them apparent, a generalization of our computation is necessary.
Moreover, it would be interesting to generalize our explicit computation
to general Calabi-Yau threefolds with some background fluxes turned
on. It would also be interesting, for  M-theory compactifications on $G_2$
manifolds, to generalize 
the computation of \cite{Dall'Agata:2005fm} to include also  non-geometric 
fluxes.


\bigskip
\noindent
\emph{Acknowledgments:} We are grateful to David Andriot, Andreas Deser, Oscar Loaiza-Brito, Dieter L\"ust,
Eran Palti,
Erik Plau\-schinn, Felix Rennecke and Christian Schmid for
discussion and Adolfo Guarino and the referee for useful comments about an earlier version
of this paper.
XG is supported by the MPG-CAS Joint Doctoral Promotion Programme.
PS is supported by a postdoctoral research fellowship 
from the Alexander von Humboldt Foundation.

\clearpage
\appendix

\section{The detailed form of the superpotentials}
\label{appendix_0}

In this appendix we present the explicit form of the superpotential
when written out in terms of the individual fluxes and moduli.

\subsubsection*{The type IIA superpotential}

Using 
\eq{
\ov{\mathfrak{H}}{}^C_{ijk}=\ov{H}_{ijk}+3\, \ov{F}^m{}_{[\underline{ij}}\,
       (-iJ_c)_{m\underline{k}]}
         &+3\, \ov{Q}_{[\underline{i}}{}^{mn} (-iJ_c)_{m\underline{j}}\,
           (-iJ_c)_{n\underline{k}]}\\[0.1cm]
    &+\ov{R}^{mnp}  (-iJ_c)_{m[\underline{i}}\, (-iJ_c)_{n\underline{j}}\,
           (-iJ_c)_{p\underline{k}]}\, ,
}
the NS-NS sector type IIA superpotential written out in terms of the fluxes
and moduli reads (similar explicit expressions have been derived in
\cite{Hassler:2013wsa})
\eq{
\label{superpotex}
    W_{\rm NS}&=
-{i\over 4} \int_X \overline{\mathfrak{H}}^C\wedge \Omega^c\\
&{1\over 4}\bigg[S\, \Big(i\ov{H}_{135} - \ov{F}^6{}_{13}\, T_3  -
           \ov{F}^2{}_{35}\, T_1 - \ov{F}^4{}_{51}\, T_2   \\
    &\hspace{1cm} -i \ov{Q}_1{}^{46} T_2 T_3 -i\ov{Q}_3{}^{62} T_1 T_3-i
            \ov{Q}_5{}^{24} T_1 T_2
+ \ov{R}^{246} T_1 T_2 T_3
                  \Big)\\
    &-U_1 \Big( i \ov{H}_{146} + \ov{F}^5{}_{14}\, T_3 -
           \ov{F}^2{}_{46}\,  T_1 + \ov{F}^3{}_{61}\,  T_2   \\
               &\hspace{1cm} -i \ov{Q}_1{}^{35}  T_2  T_3 +i \ov{Q}_4{}^{52}  T_1  T_3+i
            \ov{Q}_6{}^{23}  T_1  T_2
+ \ov{R}^{235}  T_1 T_2 T_3
                  \Big)\\
   &-U_2 \Big(i \ov{H}_{236} + \ov{F}^5{}_{23}\, T_3 +
           \ov{F}^1{}_{36}\, T_1 - \ov{F}^4{}_{62}\, T_2   \\
               &\hspace{1cm} +i \ov{Q}_2{}^{45} T_2 T_3 - i\ov{Q}_3{}^{51} T_1 T_3
                 +i \ov{Q}_6{}^{14} T_1 T_2
+ \ov{R}^{145} T_1 T_2 T_3
                  \Big)\\
   &-U_3 \Big(i \ov{H}_{245} - \ov{F}^6{}_{24}\, T_3  +
           \ov{F}^1{}_{45}\, T_1  + \ov{F}^3{}_{52}\, T_2    \\
               &\hspace{1cm} +i \ov{Q}_2{}^{36} T_2 T_3  +i \ov{Q}_4{}^{61} T_1 T_3
           -i \ov{Q}_5{}^{13} T_1 T_2
+ \ov{R}^{136} T_1 T_2 T_3
                  \Big)\bigg]\, .
}
In the R-R sector one finds
\eq{
\label{gvw2a}
   W_{\rm R}&={1\over 4}\, \int_{X} e^{iJ_c} \wedge \ov{G} \,\\[0.1cm]
  &= \frac{1}{4}\biggl[\ov{G}^{(6)}_{123456} + i \,\ov{G}^{(4)}_{3456} \, T_1 + i\, \ov{G}^{(4)}_{1256} \, T_2  + i\, \ov{G}^{(4)}_{1234}\, T_3 \\[-0.1cm]
&  \hskip1.99cm  - \ov{G}^{(2)}_{12} T_2 T_3 - 
   \ov{G}^{(2)}_{34}\, T_1 T_3 - \ov{G}^{(2)}_{56}\, T_1 T_2 - i \, T_1 T_2 T_3 \ov{G}^{(0)}\biggr]. 
}

\subsubsection*{The type IIB superpotential}

In the type IIB example, the NS-NS part of  the superpotential can be expanded as 
\eq{
W_{NS}&= -{i\over 4}\int_{X} \left( S \, \ov H+ \, \ov Q\cdot  J^c\right)\wedge \Omega_3 \\[0.1cm]
& = {1\over 4}\bigg[ {S}\, \Big(-i\ov{H}_{246} - \ov{H}_{146} U_1 -\ov{H}_{236}U_2-\ov{H}_{245}U_3 \\
  & \hspace{0.5cm}+i \, \ov{H}_{136}\, U_1\, U_2+i\, \ov{H}_{235}\, U_2\, U_3+i\, \ov{H}_{145}\, U_1\, U_3 +\ov{H}_{135}\, U_1\, U_2\, U_3
                  \Big)\\
 &+ {T_1}\Bigl(i \, \ov Q_2{}^{35} + \ov
                  Q_1{}^{35} \, U_1 - \, \ov Q_2{}^{45} \, U_2  - \ov Q_2{}^{36} \, U_3 \\
  & \hspace{0.5cm} + i\, \ov Q_1{}^{45} \, U_1 \,  U_2 - i\, \ov Q_2{}^{46}\,
                  U_2\, U_3 +  i\,\ov Q_1{}^{36} \, U_1 \, U_3 - \ov Q_1{}^{46}\, U_1 \, U_2\, U_3\Bigr) \\
    & + {T_2}\Bigl( i \, \ov Q_4{}^{51} -\, \ov Q_4{}^{52} \,
                  U_1 + \, \ov Q_3{}^{51} \, U_2  - \, \ov Q_4{}^{61} \,U_3 \\
& \hspace{0.5cm}    + i\, \ov Q_3{}^{52} \, U_1\, U_2  + i \, \ov
                  Q_3{}^{61}\, U_2 \, U_3 - i \, \ov Q_4{}^{62} \, U_1 \, U_3
                  -  \ov Q_3{}^{62}\, U_1 \, U_2 \, U_3\Bigr)\\
 & + {T_3}\Bigl(i \, \ov Q_6{}^{13} - \ov Q_6{}^{23} \, U_1 -
                  \ov Q_6{}^{14} \, U_2  + \ov Q_5{}^{13} \, U_3 \\
& \hspace{0.5cm} - i \, \ov Q_6{}^{24}\,  U_1\,  U_2  + i \, \ov Q_5{}^{14}
                  \, U_2 \, U_3 + i \, \ov Q_5{}^{23} \, U_1 \, U_3 - \ov
                  Q_5{}^{24} \, U_1 \, U_2\, U_3\Bigr)\bigg]\, 
 }
and the R-R part as
\eq{
 W_{R}&= {1\over 4} \int_X \ov{G}^{(3)} \wedge \Omega_3 \\[0.1cm]
& = {1\over 4}\bigg[\ov{G}_{246} -i \, \ov{G}_{146} \, U_1 -i \, \ov{G}_{236} \, U_2-i \, \ov{G}_{245}\, U_3  \\
 & \hspace{0.6cm} -\ov{G}_{235}\, U_2 \, U_3+\ov{G}_{145}\, U_1\,
 U_3+\ov{G}_{136}\, U_1\, U_2 + i \, \ov{G}_{135} \, U_1 \, U_2 \, U_3 \bigg]\, . 
}

\section{Details on the scalar potential}
\label{appendix_a}

The  detailed computation of the scalar potential involves
quite a number of individual terms. Recall that we compute an F-term
scalar potential $V_{F}$ from the superpotential and a K\"ahler potential
and compare the result with the dimensional reduction $V_{\rm kin}=V^{\rm
  NS}_{\rm kin}+V^{\rm R}_{\rm kin}$ of the
ten-dimensional kinetic terms. 

For the interested reader, we  provide some more details  
on the explicit computations, namely  the number of individual
terms in the scalar potential.

\subsubsection*{Counting terms for type IIA}

For the type IIA setup, we consider  four cases for which, 
in addition to all R-R fluxes, we have:
\begin{compactenum}[(i)]
\item{only $H_{ijk}$ flux turned-on}
\item{$H_{ijk}$ and geometric fluxes $F^i{}_{jk}$ turned-on}
\item{$H_{ijk}, F^i{}_{jk}$ and $Q_k{}^{ij}$ fluxes
turned-on}
\item{all fluxes turned-on}
\end{compactenum}

\noindent
Then we find the number of terms listed in table \ref{tableTypeIIA}.
\begin{table}[ht]
  \centering
  \begin{tabular}{|c||c|c|c|c||c||c|}
  \hline
   & & & & & & \\
  fluxes  & $V_{\rm kin}^{\rm NS}$ & $V_{\rm kin}^{\rm R}$& $V_{D}$&  $V_{\rm kin}-V_D$ &   $V_{F}$ &  $V_{F}-V_{\rm kin}+V_D$\\
turned-on & & & & & & (to be removed by BIs)\\
& & & & & & \\
\hline
\hline
& & & & & & \\
$H$ & 4 & 118 & 4  & 126 & 126 & 0\\
& & & & & & \\
$H, F$ & 64 & 418 & 16 & 498 & 522 & 24 $:F^2 =0$\\
& & & & & & \\
 $H, F, Q$ & 112& 1138& 28& 1446 & 1686 & 240  \\
& & & & & & $48:F^2 + H\,Q=0 $ (12 BIs)\\
 & & & & & & $48: Q^2 = 0 $ (12 BIs) \\
& & & & & & $144: F \, Q = 0$ (24 BIs)\\
& & & & & & \\
$H, F, Q, R$ & 424&1630 & 32 & 2086 & 2422 & 336\\
    & & & & & & $96:F^2 + H \, Q=0$ (12 BIs)\\
 & & & & & & $48: Q^2 + R \, F = 0$ (12 BIs) \\
 & & & & & & $192: F \, Q + H \, R = 0$ (24 BIs)\\
& & & & & & \\
    \hline
     \end{tabular} 
     \caption{Number of individual terms in the scalar potential for the type IIA orientifold. }
      \label{tableTypeIIA}
\end{table}

\noindent 
Note that the case (i) is  the standard example  in type IIA with just 
the usual NS-NS  and R-R fluxes turned on, whereas
case (ii) reproduces the results of \cite{Villadoro:2005cu} 
with the inclusion of geometric flux. 
For all cases, after imposing the Bianchi identities,  the kinetic terms 
reduce according to
\eq{ 
V_{\rm kin} = V_{\rm kin}^{\rm NS}+V_{\rm kin}^{\rm R}=V_F+V_{D}
} 
to the expected F-term potential plus a D-term, which combines with
the D-term from the localized $D6$-branes and $O6$-planes.
Using the R-R tadpole cancellation conditions, the entire D-term vanishes.

\subsubsection*{Counting terms for type IIB}

For the type IIB example, we only need to distinguish two cases
\begin{compactenum}[(i)]
\item{only $H_{ijk}$ flux turned-on}
\item{$H_{ijk}$ and non-geometric fluxes $Q_k{}^{ij}$ turned-on}
\end{compactenum}

\noindent
which are listed in table \ref{tableTypeIIB}.
\begin{table}[ht]
  \centering
  \begin{tabular}{|c||c|c|c|c||c||c||c|}
  \hline
   & & & & & & & \\
  fluxes  & $V_{\rm kin}^{\rm NS}$ & $V_{\rm kin}^{\rm R}$ & $V_{D3}$&  $V_{D7}$&  $V_{\rm kin}-V_{D}$ &   $V_{F}$ &  $V_{F}-V_{\rm kin}+V_D$\\
turned-on & & & & & & & (to be removed by BIs)  \\
& & & & & & & \\
\hline
\hline
& & & & & & & \\
 $H$ & 76 & 277 & 8 & 0  & 361 & 361 & 0\\
& & & & & & & \\
$H, Q$ & 424&1630 & 8 &24 & 2086 & 2422 & 336\\
    & & & & & & & $168: H\, Q=0$ (24 BIs)\\
 & & & & & & & $168: \, \, \, \, Q^2  = 0$ (24 BIs) \\
& & & & & & & \\
    \hline
     \end{tabular} 
     \caption{Number of individual terms in the scalar potential for the type IIB orientifold. }
      \label{tableTypeIIB}
\end{table}

It is interesting to observe that, in the most generic case when one includes
all possible fluxes, the total number of terms in the type IIA and type IIB 
case  exactly match in each column, including also the number of terms
which get nullified via the Bianchi identities.
This observation supports one of the main initial motivations for introducing 
non-geometric fluxes \cite{Shelton:2005cf}, 
namely the consistency of fluxes with T-duality.



\clearpage
\bibliographystyle{utphys}
\bibliography{references}


\end{document}